\documentclass[showkeys,preprintnumbers,amsmath,amssymb]{revtex4}

\usepackage{graphicx}
\usepackage{dcolumn}
\usepackage{bm}
\begin{document}

\title{A case of entanglement generation between causally disconnected atoms}

\author{Juan Le\'on}
\email{leon@imaff.cfmac.csic.es} \homepage{http://www.imaff.csic.es/pcc/QUINFOG/}

\author{Carlos Sab\'{i}n}%
 \email{csl@imaff.cfmac.csic.es}
\homepage{http://www.imaff.csic.es/pcc/QUINFOG/}
\affiliation{%
Instituto de F\'{i}sica Fundamental, CSIC
 \\
Serrano 113-B, 28006 Madrid, Spain.\\
}%


\date{\today}

\begin{abstract}
We analyze the entanglement generated in a finite time between a pair of space-like separated atoms, one of which emits a
photon. As we show to order $e^2$, the origin of entanglement can be traced back to the uncertainty about which one of the atoms
emitted the photon. We check this by comparing the time behaviors of the emission processes allowed by energy conservation vs.
those forbidden by the same reason. No physical signal propagates between the atoms in the processes considered, however an
effective light cone separating non-entangled from entangled regions in space-time emerges from our calculations.
\end{abstract}

\maketitle


Quantum Mechanics may predict correlations between space-like separated systems. These may violate Bell inequalities too
\cite{bell,aspect}, in which case they can not be  explained by strategies arranged in the past \cite{gisin}. In this work we
show the generation of quantum correlations between two space-like separated parties which do not exchange any signal or any
other physical interaction. We will consider a pair of neutral two level atoms $A$ and $B$ separated by a distance $L$
\cite{fermi,hegerfeldt,franson,reznik,reznikII,fransonII}.  We study perturbatively, to order $\alpha$, their local interactions
with the electromagnetic field during a finite time $T$, and compute the correlations in the final state of the atoms
\cite{cabrillo,lamata} when a lone photon is produced during that time. The bi-atom state shows a finite concurrence that we
compute in terms of $L$ and $T$. Our results show the emergence of an effective non-signaling, which is remarkable in a case
like this, where nothing at all is exchanged between the atoms. On the one side, there is no phenomenon to whom trace back the
change of behavior of the concurrence at $L = cT$. On the other, $T$ is the interaction duration, not the time employed by any
propagating signal.

We begin considering that the field is initially  in the vacuum state, including  in the final state the cases with 0, 1 and 2
photons, to analyze perturbatively the amplitudes to order $\alpha$. We assume that the wavelengths relevant in the interaction
with the atoms, and the separation between them, are much longer than the atomic dimensions.  The dipole approximation,
appropriate to these conditions,  permits the splitting of the system Hamiltonian into two parts $H = H_0 + H_I$ that are
separately gauge invariant. The first part is the Hamiltonian in the absence of interactions other than the potentials that keep
$A$ and $B$ stable, $H_0 = H_A + H_B + H_{\mbox{field}}$. The second contains all  the interaction of the atoms with the field
$H_I = - \frac{1}{\epsilon_0}\sum_{N=A,B} \mathbf{d}_N(\mathbf{x}_N,t)\,\mathbf{D}(\mathbf{x}_N,t)$,where $\mathbf{D}$ is the
electric displacement field, and $\mathbf{d}_N \,=\,\sum_i\, e\,\int d^3 \mathbf{x}_i\,
\langle\,E_N\,|\,(\mathbf{x}_i-\mathbf{x}_N)\,|\,G_N\,\rangle$ is the electric dipole moment of atom $N$, that we will take as
real and of equal magnitude for both atoms.

 In what follows we choose a system  given initially by the product state $|\,\psi\,\rangle_0\,=\,  |\,E\,G\,\rangle\cdot|\,0\,\rangle$, in which
atom $A$ is in the excited state $|\,E\,\rangle$, atom $B$ in the ground state $|\,G\,\rangle$, and the field in the vacuum
state $|\,0\,\rangle$. The system then evolves under the effect of the interaction during a lapse of time $T$ into a state that,
to  order $\alpha$, can be given in the interaction picture as
\begin{eqnarray}
|\mbox{atom}_1,\mbox{atom}_2,\mbox{field}\rangle_{T} =  ((1+a)\,|\,E\,G\rangle + b\,|\,G\,E\rangle)\,|\,0\rangle\nonumber\\
 +(u\,|\,G\,G\,\rangle+ v\,|\,E\,E\,\rangle)\,|\,1\,\rangle+
(f\,|\,E\,G\rangle+ g\,|\,G\,E\rangle)\,|\,2\rangle\  \label{a}
\end{eqnarray}
where
\begin{eqnarray}
a&=&-\frac{1}{2}\langle0|T(\mathcal{S}_A \mathcal{S}_A+\mathcal{S}_B\mathcal{S}_B)|0\rangle,\, b= -\langle0|T(\mathcal{S}^+_B
\mathcal{S}^-_A)|0\rangle\nonumber\\
u\,&=&\,-i\,\langle\,1\,|\, \mathcal{S}^-_A\,|\,0\,\rangle,\, v\,=\,-i\,\langle\,1\,|\,
\mathcal{S}^+_B\,|\,0\,\rangle \label{b}\\
f&=&-\frac{1}{2}\langle2|T(\mathcal{S}_A \mathcal{S}_A+\mathcal{S}_B\mathcal{S}_B)|0\rangle,\, g=-\langle2|T(\mathcal{S}^+_B
\mathcal{S}^-_A)|0\rangle\nonumber
\end{eqnarray}
and $|\,n\,\rangle,\,\, n=\,0,\,1,\,2$ is a shorthand for the state of $n$ photons with definite momenta and polarizations, i.e.
$|\,1\,\rangle\,=\,|\mathbf{k},\, \mathbf{\epsilon}\,\rangle$, etc. Notice that among all the terms that contribute to the final
state (\ref{a}) only $b$ corresponds to interaction between both atoms. This would change at higher order in $\alpha$. Here, $a$
describes intra-atomic radiative corrections,   $u$ and $v$ single photon emission by one atom, and $g$ by both atoms, while $f$
corresponds to two photon emission by a single atom.

Finally, in the dipole approximation the actions $\hbar\, \mathcal{S}^{\pm}_N$ in (\ref{b}) reduce to
\begin{eqnarray}
\mathcal{S}^{\pm}_N\,=\,- \frac{1}{\hbar}  \int_0^T\, dt \: e^{\pm i\Omega t}\,
\mathbf{d}_N\,\mathbf{E}(\mathbf{x}_N,t)\label{c}
\end{eqnarray}
where  $\Omega = \omega_E -\omega_G$ is the transition frequency, and we are neglecting atomic recoil. This depends on the
atomic properties $\Omega$ and $\mathbf{d}$, and on the interaction time $T$. In our calculations we will take $(\Omega
|\mathbf{d}|/e c) = 5\,\cdot 10^{-3}$, which is of the same order as the 1s $\rightarrow$ 2p transition in the hydrogen atom,
consider $\Omega T \gtrsim 1$, and analyze the cases $(L/cT)\simeq 1$ around the light cone. The effective coupling, given by
the ratio $(|\mathbf{d}|/e L) \simeq \vartheta(10^{-3})$ here, could be larger if $\Omega T < 1$ entering into the Zeno region
(incidentally, the only atom atom interaction $|b|\propto T^4$ for very small $T$ as shown in Ref. \onlinecite{thiru}, not to
$T^2$ as is sometimes stated).

Given a definite field state $|\,n\,\rangle$ the pair of atoms is in a pure two qubits state as shown in (\ref{a}). We will
denote these states by $|\,A,B,n\,\rangle$,  $\rho_{AB}^{(n)}\,=\,|A,B,n\rangle\,\langle A,B,n\,|$, and $\rho_{A}^{(n)}\,=\,Tr_B
\,\rho_{AB}^{(n)}$ in the following, and will compute the entropy of entanglement $\mathbb{S}^{(n)}$ \cite{bennett} and  the
concurrence $\mathbb{C}^{(n)}$  \cite{wootters} for them. Our computations will be done for the illustrative case where both
dipoles are parallel and orthogonal to the line joining $A$ and $B$. This geometrical configuration would correspond to an
experimental set up in which the dipoles are induced by suitable external fields.

We first consider that the field state is not detected, that is, we trace over the field degrees of freedom. Then the atomic
state is represented by the following density matrix (in the basis $\{|EE\rangle,|EG\rangle,|GE\rangle,|GG\rangle\}$):
\begin{eqnarray}
\rho_{AB}=\left( \begin{array}{c c c c}|v|^2&0&0&vu^*\\0&|1+a|^2+|f|^2&(1+a)b^*+fg^*&0\\0& b(1+a)^*+f^*g&|b|^2+|g|^2&0\\
v^*u&0&0&|u|^2
\end{array}\right)N^{-1}
\end{eqnarray}
where $N=|1+a|^2 +|b|^2+|u|^2+|v|^2+|f|^2+|g|^2.$
\begin{figure}[h]
\includegraphics[width=0.90\textwidth]{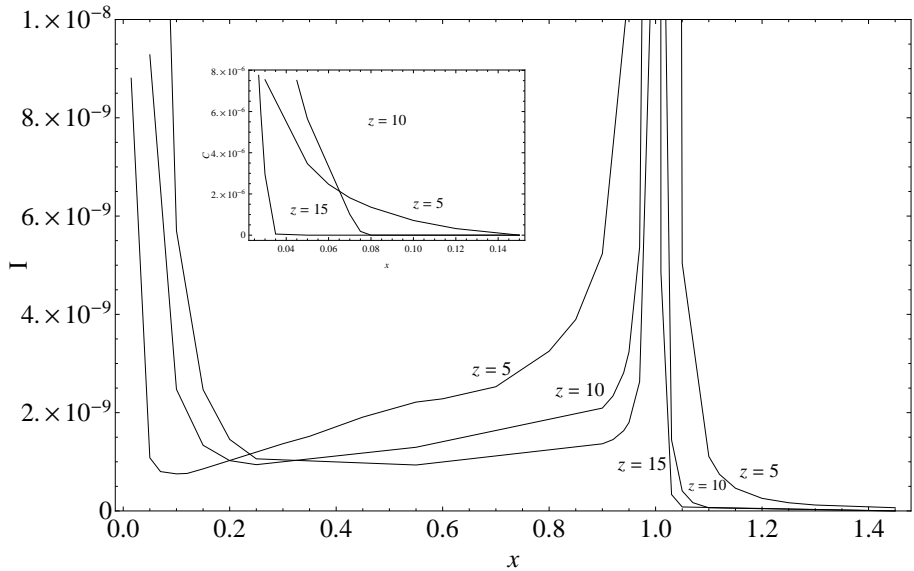}
\caption{Mutual information $\mathbb{I}(\rho_{AB})$ as a function of $x=L/cT$ for three values of $z=\Omega L/c$. The inset
shows the finite concurrences that are possible only for small values of $x$.}
\end{figure}
It can be shown that the concurrence associated to this density always vanishes except for a bounded range of small values of
$x=L/cT$. In the inset in Fig. 1 we show the concurrence in this region for several values of $z=\Omega L/c$. It grows
asymptotically as $x \rightarrow 0$,{\it i.e.} when $T\rightarrow \infty$. Out of this region $\rho_{AB}$ is a separable state
with no quantum correlations, either inside or outside the light cone. The mutual information $\mathbb{I}(\rho_{AB})=
\mathbb{S}(\rho_{A})+\mathbb{S}(\rho_{B})-\mathbb{S}(\rho_{AB})$, which measures the total correlations between both parties, is
completely classical in this case. We show this quantity in Fig. 1 for different values of $z$. In Ref. \onlinecite{conjuan} we
showed how the concurrence becomes finite (and the correlations quantum) if specific field states , with $n=0, 1$ or $2$, are
considered.

In what follows we shall focus on the $n=1$ case where the atoms excite one photon from the vacuum,  jumping to the state
$(\,u\,|\,G\,G\,\rangle\,+\,v\,|\,E\,E\,\rangle)/c_1$, (with $c_1 = \sqrt{|\,u\,|^2\,+\,|\,v\,|^2}$), during the time interval
$T$. The concurrence is
\begin{equation}
\mathbb{C}^{(1)} \,=\,2 |\,l\,|/c_1^2\ ,
 \end{equation}
where $l=v\,u^* = Tr_1\,\langle\,1\,|\, \mathcal{S}^+_B\,|\,0\,\rangle\,\langle\,1\,|\,
\mathcal{S}^-_A\,|\,0\,\rangle^*\,=\,\langle\,0\,|\, \mathcal{S}^+_A\,\mathcal{S}^+_B\,|\,0\,\rangle $. So, even if this case
only describes independent local phenomena attached to the emission of one photon by either atom $A$ or $B$, the concurrence
comes from the tangling between the amplitudes $u$ and $v$ which have different loci. The state of the photon emitted by $A$ and
the state of $A$ are correlated in the same way as the state of the photon emitted by $B$ with the state of $B$ are. These
independent field-atom correlations are transferred to atom-atom correlations when we trace out a photon line with different
ends, $A$ and $B$, when computing $v\,u^*$. In fact, while $|u|^2$ and $|v|^2$ are independent of the distance $L$ between the
atoms,
\begin{equation}
 l= - {c d_A^id_B^j \over \hbar \epsilon_0}\, \{(\delta_{ij}-\hat{L}_i \hat{L}_j)M''(L)+(\delta_{ij}+\hat{L}_i \hat{L}_j){ M'(L)\over L}
  \}
 \end{equation}
where

\begin{equation}
 M(L) =
 \int_{0}^{\infty}dk \,{\sin{k L}\over L}\,\delta^T(\Omega+c k)\,\delta^T(\Omega-c k)\label{g}
\end{equation}
which depends explicitly on $L$. Above we used $\delta^T(\omega)\,=\, \sin (\,\omega\,T/2)/(\pi \omega)$, which becomes
$\delta(\omega)$ in the limit $T\rightarrow \infty$. In Fig. 2 we represent $\mathbb{C}^{(1)}$ in front of $x={L/cT}$ for some
values of $z=\Omega L/c$. As the Figure shows, there may be  a significative amount of concurrence for all finite $x$,
indicating that $\rho^{(1)}$ is an entangled state inside and outside the light cone. The peak at $x=1$ comes from the term with
phase $k(L - c T)$ that can be singled out from the linear combination of phasors in the integrand of (\ref{g}).
\begin{figure}[h]
\includegraphics[width=0.9\textwidth]{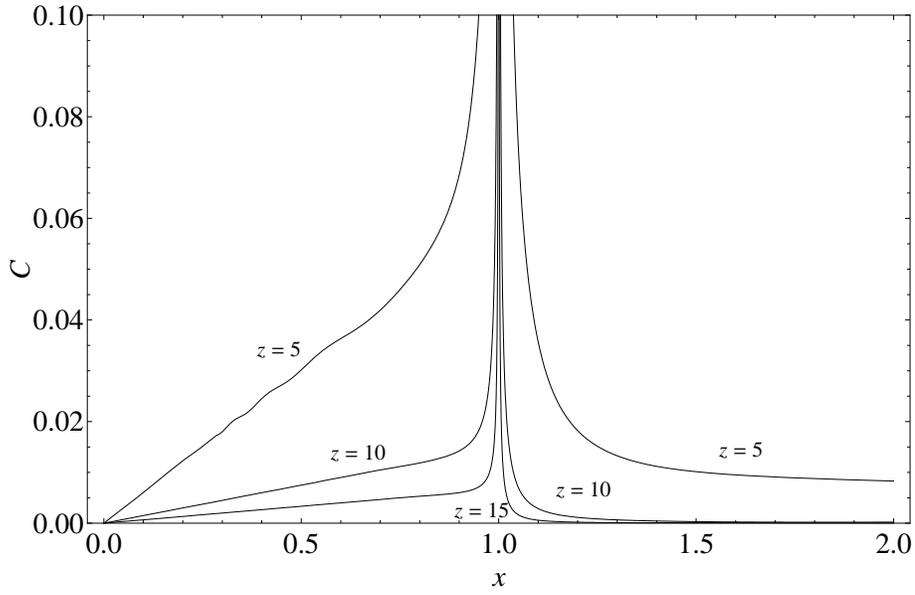}
\caption{Concurrence for one photon final state  as a function of $x={L/cT}$ for three values of $z=\Omega L/c$ when the initial
atomic  state was $|E\,G\,\rangle$.}
\end{figure}

Here we have a lone photon whose source we can not tell. It might be $A$ or $B$, with the values of $l$ and $\mathbb{C}^{(1)}$
depending on their indistinguishability. Eventually, conservation of energy will forbid the process $G \rightarrow E + \gamma$
for large interaction times. Therefore, $v$, $l$ and $\mathbb{C}^{(1)}$ will vanish as $T$ grows to infinity, as can be deduced
from the vanishing of $\delta^T(\Omega+c k)$ for $T\rightarrow \infty$.

The entropy of entanglement gives an alternative description of the situation. Its computation requires tracing over one of the
parts $A$ or $B$, so no information is left in $\mathbb{S}^{(1)}$ about $L$, but it still gives information about the relative
contribution of both participating states $|E\,E\,\rangle$ and $|G\,G\,\rangle$  to the final state.  In terms of  $\eta_1\,=\,
|\,v\,|^2/{c_1}^2 \in (0,1)$, we have
 \begin{equation}
\mathbb{S}^{(1)}\,=\, - (1-\eta_1)\, \log (1-\eta_1)\,-\,\eta_1\,\log \eta_1 \label{e}
\end{equation}

Would not be for the difference between $\Omega + ck$ and  $\Omega -ck$, $v$ should be equal to $u$, then $\eta_1\,= 0.5$, and
$\mathbb{S}^{(1)}$ would attain its maximum value. Not only this is not the case but, as said above, $v$ will vanish with time
and only $|G\,G\,\rangle$ will be in the final asymptotic state. Notice the result, indistinguishability was swept away because
for large $T$  we know which atom ($A$) emitted the photon. Therefore, the entropy will eventually vanish for large interaction
times.
\begin{figure}[b]
\includegraphics[width=0.9\textwidth]{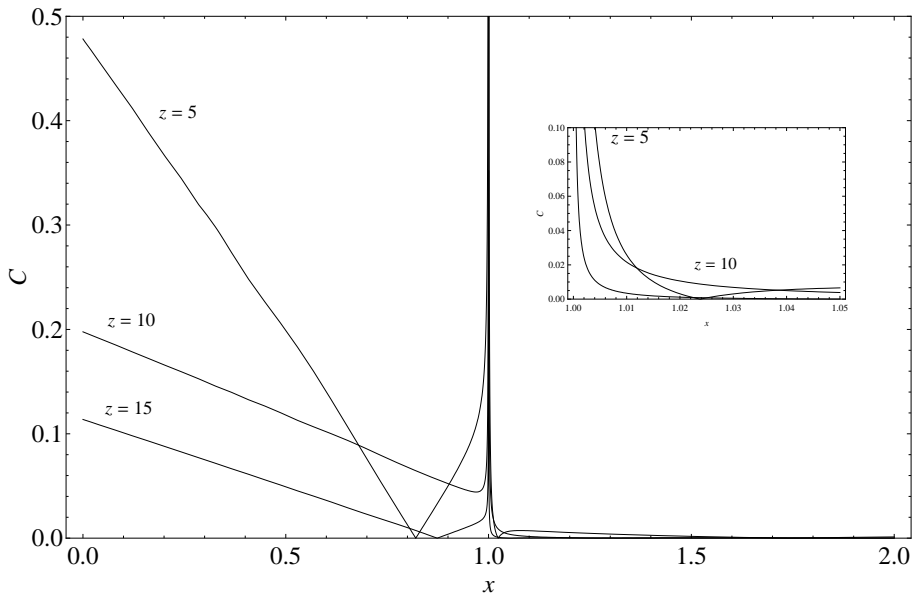}
\caption{Concurrence for one photon final state if $|E\,E\,\rangle$ is the initial state as a function of $x={L/cT}$ for three
values of $z=\Omega L/c$. The values of $\mathbb{C}$ for $x>1$ are of the same order as those displayed in Fig. 2.}
\end{figure}
This is not the case if the initial atomic state is $|E\,E\,\rangle$ or $|G\,G\,\rangle$. In the first case, the final one
photon atomic state would be $(\,u_{A}\,|\,G\,E\,\rangle\,+\,u_{B}\,|\,E\,G\,\rangle)/c'_1$, where $u_A$ is the same function of
$\mathbf{x}_A$ as $u_B$ is of $\mathbf{x}_B$, and  $c'_1 = \sqrt{2|\,u\,|^2}$ . Now, indistinguishability persists for large $T$
and so does entropy and concurrence. In particular, entropy attain its maximum value, as commented above. The $L$-dependent
concurrence is $\mathbb{C}^{(1)} \,=\,2 |\,l'\,|/(c'_1)^2$ , where $l'=\,\langle\,0\,|\,
\mathcal{S}^+_A\,\mathcal{S}^-_B\,|\,0\,\rangle$. We represent it in Fig. 3. In the second case, the atoms would be in a state
given by $(\,v_{A}\,|\,E\,G\,\rangle\,+\,v_{B}\,|\,G\,E\,\rangle)/c''_1$, being $c''_1 = \sqrt{2|\,v\,|^2}$. The difference is
that now both processes are eventually forbidden by energy conservation at large $T$. Entropy, which achieve again its maximum
value, is not able to detect this difference because both terms contribute the same (a vanishing amplitude) to the state. The
concurrence is now $\mathbb{C}^{(1)} \,=\,2 |\,l''\,|/(c''_1)^2$  with $l''=\,\langle\,0\,|\,
\mathcal{S}^-_A\,\mathcal{S}^+_B\,|\,0\,\rangle$. Indistinguishability, represented by $l''$ persists for large $T$, but for a
physical situation  whose probability  vanishes. Finally, in Fig. 4 we have represented the concurrence for the case where the
inial atomic state was $|E\,E\,\rangle$ in terms of the inter-atomic distance for three fixed values of time. What we obtain is
a shift of the concurrence features to longer $L$ as $T$ grows (so that they appear at the same $(L/cT)$), in such a way that,
even if  $T$ is just the duration of the interaction, it plays the role of propagation time for the generated correlations. They
are negligible small for large $L$, peak at the ``light cone" but, on the other hand grow, as we would expect, for larger
interaction times.

\begin{figure}[h]
\includegraphics[width=0.9\textwidth]{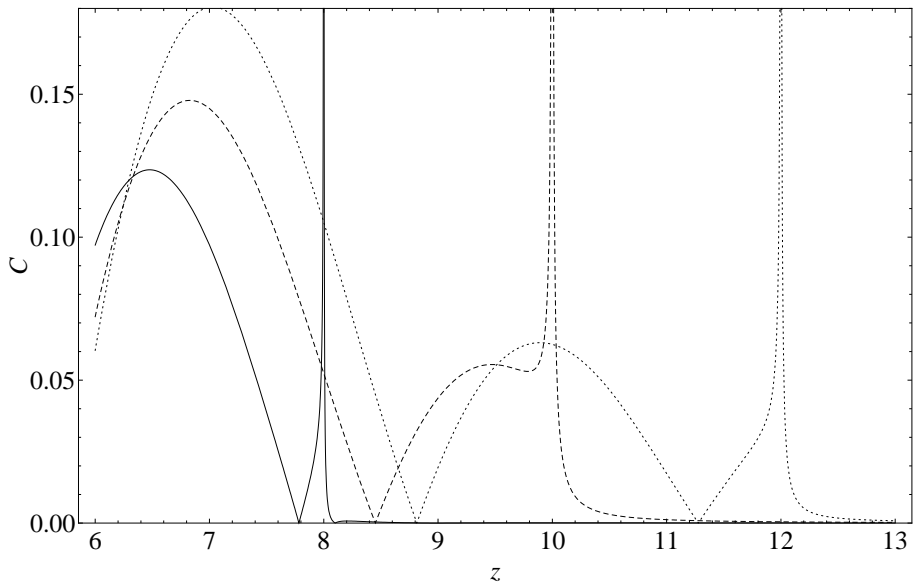}
\caption{Concurrence for one photon final state if $|E\,E\,\rangle$ is the initial state as a function of $z=\Omega L/c$ for
three representative values of the time $\Omega T= 8, 10, 12$ with peaks at $z=8, 10$ and $12$ respectively.}
\end{figure}

This work was supported by Spanish MEC FIS2005-05304 and CSIC 2004 5 OE 271 projects. C.S. acknowledges support from CSIC I3
program.


\begin{thebibliography}{15}
\expandafter\ifx\csname natexlab\endcsname\relax\def\natexlab#1{#1}\fi \expandafter\ifx\csname bibnamefont\endcsname\relax
  \def\bibnamefont#1{#1}\fi
\expandafter\ifx\csname bibfnamefont\endcsname\relax
  \def\bibfnamefont#1{#1}\fi
\expandafter\ifx\csname citenamefont\endcsname\relax
  \def\citenamefont#1{#1}\fi
\expandafter\ifx\csname url\endcsname\relax
  \def\url#1{\texttt{#1}}\fi
\expandafter\ifx\csname urlprefix\endcsname\relax\def\urlprefix{URL }\fi \providecommand{\bibinfo}[2]{#2}
\providecommand{\eprint}[2][]{\url{#2}}

\bibitem[{\citenamefont{Bell}(1954)}]{bell}
\bibinfo{author}{\bibfnamefont{J.~S.} \bibnamefont{Bell}},
  \bibinfo{journal}{Physics} \textbf{\bibinfo{volume}{1}}, \bibinfo{pages}{195}
  (\bibinfo{year}{1954}).

\bibitem[{\citenamefont{Aspect}(1999)}]{aspect}
\bibinfo{author}{\bibfnamefont{A.}~\bibnamefont{Aspect}},
  \bibinfo{journal}{Nature} \textbf{\bibinfo{volume}{398}},
  \bibinfo{pages}{189} (\bibinfo{year}{1999}).

\bibitem[{\citenamefont{Masanes et~al.}(2006)\citenamefont{Masanes, Ac{\'i}n,
  and Gisin}}]{gisin}
\bibinfo{author}{\bibfnamefont{L.}~\bibnamefont{Masanes}},
  \bibinfo{author}{\bibfnamefont{A.}~\bibnamefont{Ac{\'i}n}}, \bibnamefont{and}
  \bibinfo{author}{\bibfnamefont{N.}~\bibnamefont{Gisin}},
  \bibinfo{journal}{Phys. Rev. A} \textbf{\bibinfo{volume}{73}},
  \bibinfo{pages}{012112} (\bibinfo{year}{2006}).

\bibitem[{\citenamefont{Fermi}(1932)}]{fermi}
\bibinfo{author}{\bibfnamefont{E.}~\bibnamefont{Fermi}},
  \bibinfo{journal}{Rev.\ Mod.\ Phys.} \textbf{\bibinfo{volume}{4}},
  \bibinfo{pages}{87} (\bibinfo{year}{1932}).

\bibitem[{\citenamefont{Hegferfeldt}(1998)}]{hegerfeldt}
\bibinfo{author}{\bibfnamefont{G.~C.} \bibnamefont{Hegferfeldt}},
  \bibinfo{journal}{Annalen Phys.} \textbf{\bibinfo{volume}{7}},
  \bibinfo{pages}{716} (\bibinfo{year}{1998}).

\bibitem[{\citenamefont{Franson}()}]{franson}
\bibinfo{author}{\bibfnamefont{J.~D.} \bibnamefont{Franson}},
  \eprint{quant-ph/0704.1468}.

\bibitem[{\citenamefont{Reznik et~al.}(2005)\citenamefont{Reznik, Retzker, and
  Silman}}]{reznik}
\bibinfo{author}{\bibfnamefont{B.}~\bibnamefont{Reznik}},
  \bibinfo{author}{\bibfnamefont{A.}~\bibnamefont{Retzker}}, \bibnamefont{and}
  \bibinfo{author}{\bibfnamefont{J.}~\bibnamefont{Silman}},
  \bibinfo{journal}{Phys.\ Rev.\ A} \textbf{\bibinfo{volume}{71}},
  \bibinfo{pages}{042104} (\bibinfo{year}{2005}).

\bibitem[{\citenamefont{Retzker et~al.}(2005)\citenamefont{Retzker, Cirac, and
  Reznik}}]{reznikII}
\bibinfo{author}{\bibfnamefont{A.}~\bibnamefont{Retzker}},
  \bibinfo{author}{\bibfnamefont{J.~I.} \bibnamefont{Cirac}}, \bibnamefont{and}
  \bibinfo{author}{\bibfnamefont{B.}~\bibnamefont{Reznik}},
  \bibinfo{journal}{Phys.\ Rev.\ Lett.} \textbf{\bibinfo{volume}{94}},
  \bibinfo{pages}{050504} (\bibinfo{year}{2005}).

\bibitem[{\citenamefont{Franson and Donegan}(2002)}]{fransonII}
\bibinfo{author}{\bibfnamefont{J.~D.} \bibnamefont{Franson}} \bibnamefont{and}
  \bibinfo{author}{\bibfnamefont{M.~M.} \bibnamefont{Donegan}},
  \bibinfo{journal}{Phys.\ Rev. A} \textbf{\bibinfo{volume}{65}},
  \bibinfo{pages}{052107} (\bibinfo{year}{2002}).

\bibitem[{\citenamefont{Cabrillo et~al.}(1999)\citenamefont{Cabrillo, Cirac,
  P.Garc{\'i}a-Fern{\'a}ndez, and Zoller}}]{cabrillo}
\bibinfo{author}{\bibfnamefont{C.}~\bibnamefont{Cabrillo}},
  \bibinfo{author}{\bibfnamefont{J.~I.} \bibnamefont{Cirac}},
  \bibinfo{author}{\bibnamefont{P.Garc{\'i}a-Fern{\'a}ndez}}, \bibnamefont{and}
  \bibinfo{author}{\bibfnamefont{P.}~\bibnamefont{Zoller}},
  \bibinfo{journal}{Phys.\ Rev.\ A} \textbf{\bibinfo{volume}{59}},
  \bibinfo{pages}{001025} (\bibinfo{year}{1999}).

\bibitem[{\citenamefont{Lamata et~al.}(2007)\citenamefont{Lamata,
  Garc{\'i}a-Ripoll, and Cirac}}]{lamata}
\bibinfo{author}{\bibfnamefont{L.}~\bibnamefont{Lamata}},
  \bibinfo{author}{\bibfnamefont{J.~J.} \bibnamefont{Garc{\'i}a-Ripoll}},
  \bibnamefont{and} \bibinfo{author}{\bibfnamefont{J.~I.} \bibnamefont{Cirac}},
  \bibinfo{journal}{Phys.\ Rev.\ Lett.} \textbf{\bibinfo{volume}{98}},
  \bibinfo{pages}{010502} (\bibinfo{year}{2007}).

\bibitem[{\citenamefont{Craig and Thirunamachandran}(1992)}]{thiru}
\bibinfo{author}{\bibfnamefont{D.~P.} \bibnamefont{Craig}} \bibnamefont{and}
  \bibinfo{author}{\bibfnamefont{T.}~\bibnamefont{Thirunamachandran}},
  \bibinfo{journal}{Chem.\ Phys.} \textbf{\bibinfo{volume}{167}},
  \bibinfo{pages}{229} (\bibinfo{year}{1992}).

\bibitem[{\citenamefont{Bennett et~al.}(1996)\citenamefont{Bennett, Bernstein,
  Popescu, and Schumacher}}]{bennett}
\bibinfo{author}{\bibfnamefont{C.~H.} \bibnamefont{Bennett}},
  \bibinfo{author}{\bibfnamefont{H.~J.} \bibnamefont{Bernstein}},
  \bibinfo{author}{\bibfnamefont{S.}~\bibnamefont{Popescu}}, \bibnamefont{and}
  \bibinfo{author}{\bibfnamefont{B.}~\bibnamefont{Schumacher}},
  \bibinfo{journal}{Phys.\ Rev.\ A} \textbf{\bibinfo{volume}{53}},
  \bibinfo{pages}{2046} (\bibinfo{year}{1996}).

\bibitem[{\citenamefont{Hill and Wootters}(1997)}]{wootters}
\bibinfo{author}{\bibfnamefont{S.}~\bibnamefont{Hill}} \bibnamefont{and}
  \bibinfo{author}{\bibfnamefont{W.~K.} \bibnamefont{Wootters}},
  \bibinfo{journal}{Phys.\ Rev.\ Lett.} \textbf{\bibinfo{volume}{78}},
  \bibinfo{pages}{5022} (\bibinfo{year}{1997}).

\bibitem[{\citenamefont{Le{\'o}n and Sab{\'i}n}()}]{conjuan}
\bibinfo{author}{\bibfnamefont{J.}~\bibnamefont{Le{\'o}n}} \bibnamefont{and}
  \bibinfo{author}{\bibfnamefont{C.}~\bibnamefont{Sab{\'i}n}},
  \emph{\bibinfo{title}{Generation of atom-atom correlations around the light
  cone}}, \eprint{arXiv:0804.4641}.

\end{thebibliography}
\end{document}